\documentclass[epj-spec]{svjour}

\usepackage{graphics}

\begin{document}

\title{An Optical Lattice Clock with Spin-polarized $^{87}$Sr Atoms}
\author{Xavier Baillard\inst{1} \and Mathilde Fouch\'{e}\inst{1}\fnmsep\inst{3} \and Rodolphe Le Targat\inst{1}
\and Philip G. Westergaard\inst{1} \and Arnaud Lecallier\inst{1} \and Fr\'{e}d\'{e}ric Chapelet\inst{1} \and Michel
Abgrall\inst{1} \and Giovanni D. Rovera\inst{1} \and Philippe Laurent\inst{1} \and Peter Rosenbusch\inst{1} \and S\'{e}bastien
Bize\inst{1} \and Giorgio Santarelli\inst{1} \and Andr\'{e} Clairon\inst{1} \and Pierre
Lemonde\inst{1}\fnmsep\thanks{\email{pierre.lemonde@obspm.fr}} \and Gesine Grosche\inst{2} \and Burghard Lipphardt\inst{2} \and
Harald Schnatz\inst{2}}

\institute{LNE-SYRTE, Observatoire de Paris, 61, avenue de l'Observatoire, 75014, Paris, France. \and Physikalisch-Technische
Bundesanstalt, Bundesallee 100, 38116 Braunschweig, Germany. \and Laboratoire Collisions Agr\'{e}gats R\'{e}activit\'{e}, UMR
5589 CNRS - Universit\'{e} Paul Sabatier Toulouse 3, IRSAMC, 31062 Toulouse cedex 9, France.}

\abstract{ We present a new evaluation of an $^{87}$Sr optical lattice clock using spin polarized atoms. The frequency of the
$^1S_0\rightarrow\,^3P_0$ clock transition is found to be 429\,228\,004\,229\,873.6\,Hz with a fractional accuracy of $2.6\times
10^{-15}$, a value that is comparable to the frequency difference between the various primary standards throughout the world.
This measurement is in excellent agreement with a previous one of similar accuracy\,\cite{Boyd07}.}

\maketitle

\section{Introduction} \label{sec:intro}

The possibility to build high accuracy optical clocks with $^{87}$Sr atoms confined in an optical lattice is now well
established. Since this idea was published\,\cite{KatoPal03}, experiments rapidly proved the possibility to obtain narrower and
narrower resonances with atoms in the Lamb-Dicke regime\,\cite{Takamoto03,Ludlow06,Brusch06}. The narrowest observed resonances
now have a width in the Hz range\,\cite{Boyd06} and the corresponding potential fractional frequency instabilities are better
than $10^{-16}$ over 1\,s of averaging time. On the other hand, systematic effects were also shown to be highly controllable. It
was theoretically demonstrated that the residual effects of the atomic motion could be reduced down to the $10^{-18}$ level for a
lattice depth as small as $10\,E_r$\,\cite{Lemonde05}, with $E_r$ the recoil energy associated with the absorption or emission of
a lattice photon. Higher order frequency shifts due to the trapping light were then also shown to be controllable at that
level\,\cite{Brusch06}\footnote{The first order shift can be made to vanish in this type of clocks at the so-called "magic
wavelength".}. Altogether, the accuracy of the frequency measurement of the $^1S_0-{^3P_0}$ clock transition of Sr has steadily
improved by four orders of magnitude since its first direct measurement in 2003\,\cite{Courtillot03}. Three independent
measurements performed in Tokyo university\,\cite{Takamoto06}, JILA\,\cite{Ludlow06} and SYRTE\,\cite{Letargat06} were reported
with a fractional uncertainty of $10^{-14}$ giving excellent agreement. Recently, the JILA group improved their uncertainty down
to $2.5\times 10^{-15}$\,\cite{Boyd07}, a value that is comparable to the frequency difference between the various primary
standards throughout the world\,\cite{Wolf063}. We report here a new and independent measurement of this clock transition with an
accuracy of $2.6\times 10^{-15}$. The major modification as compared to our previous evaluation is an improved control of the
Zeeman effect. By applying a bias field of typically $0.1$\,mT and pumping atoms into extreme Zeeman states (we alternate
measurements with $m_F=+9/2$ and $m_F=-9/2$) we cancel the first order Zeeman effect while getting a real time measurement of the
actual magnetic field seen by the atoms\,\cite{Takamoto06}. The measured frequency of the $^1S_0\rightarrow{^3P_0}$ clock
transition of $^{87}$Sr is 429 228 004 229 873.6\,(1.1)\,Hz. This value differs from the one of Ref.\,\cite{Boyd07} by 0.4\,Hz
only.

\section{Experimental setup}
\subsection{Atom manipulation} \label{sec:setup}
\begin{figure}
\begin{center}
\resizebox{0.75\columnwidth}{!}{
\includegraphics{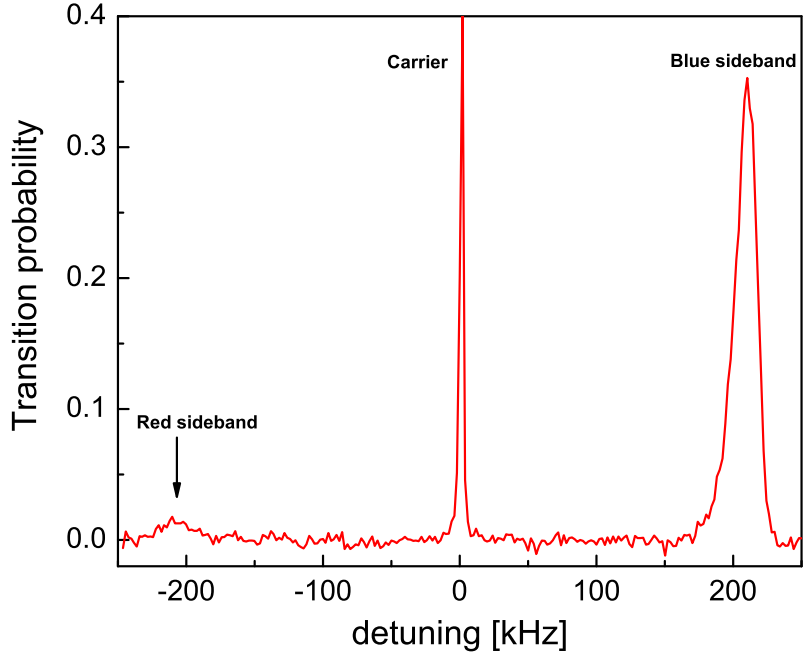} }
\caption{Spectrum at high power of the carrier and the first two longitudinal sidebands of the trapped atoms. The ratio between
both sidebands is related to the population of the ground state of the trap. 95\,\% of the atoms are in the lowest vibrational
state of the lattice wells.} \label{fig:sidebands}
\end{center}
\end{figure}

The apparatus is derived from the one described in Ref.\,\cite{Letargat06}. The clock is operated sequentially, with a typical
cycle duration of 400\,ms. We use a dipole trap formed by a vertical standing wave at 813.428\,nm inside an enhancement
Fabry-P\'{e}rot cavity. The depth of the wells of the resulting lattice is typically 100\,$\mu$K. A beam of $^{87}$Sr atoms from
an oven at a temperature of about 450\,$^{\circ}$C is sent through a Zeeman slower and then loaded into a magneto-optical trap
(MOT) based on the $^1S_0\rightarrow\,^1P_1$ transition at 461\,nm. The MOT temperature is about 1\,mK. The dipole trap laser is
aligned in order to cross the center of the MOT. Two additional laser beams tuned to the $^1S_0\rightarrow\,^3P_1$ and
$^3P_1\rightarrow\,^3S_1$ transitions, at 689\,nm and 688\,nm respectively, are superimposed on the trapping laser. The atoms
that cross these beams are therefore drained into the metastable states $^3P_0$ and $^3P_2$ at the center of the trap, and those
with a small enough kinetic energy remain confined in the potential wells forming the lattice. The MOT and drain lasers are then
switched off and atoms are optically pumped back to the ground state, where they are further cooled using the narrow
$^1S_0\rightarrow\,^3P_1$ transition. About 95\,\% of the atoms are cooled down to the ground state of the trap (see
Fig.\,\ref{fig:sidebands}), corresponding to a temperature of 3\,$\mu$K. They are optically pumped into either the ($^1S_0$,
$m_F=9/2$) or ($^1S_0$, $m_F=-9/2$) Zeeman sub-state. This is achieved by means of a bias magnetic field of about $=10^{-4}$\,T
and a circularly polarized laser ($\sigma^+$ or $\sigma^-$ depending on the desired $m_F$ state) tuned to the
$^1S_0(F=9/2)\rightarrow\,^3P_1(F=9/2)$ transition. This transition is power-broadened to a few hundreds of kHz. The magnetic
field can then be switched to a different value (up to a fraction of a mT) for the clock transition interrogation. We use a
$\pi$-polarized laser at 698\,nm to probe the $^1S_0\rightarrow\,^3P_0$ transition with adjustable frequency to match the desired
($m_F=\pm9/2\rightarrow m_F=\pm9/2$) transition. Finally the populations of the two states $^1S_0$ and $^3P_0$ are measured by
laser induced fluorescence using two blue pulses at 461\,nm separated by a repumping pulse.

\subsection{Measurement scheme} \label{sec:meas}

The spectroscopy of the clock transition is performed with an extended-cavity diode laser at 698\,nm which is pre-stabilized with
an interference filter\,\cite{Baillard06}. The laser is stabilized to an ultra-stable cavity of finesse $F=25000$, and its
frequency is constantly measured by means of a femtosecond fiber laser \cite{Tamura93,Kubina05} referenced to the three atomic
fountain clocks FO1, FO2 and FOM. The femtosecond fiber laser is described in paragraph \ref{sec:femto}.

The fountain ensemble used in this measurement is described extensively in \cite{Bize04,Bize05}. The three atomic fountains FO1,
FO2 and FOM are used as primary frequency standards measuring the frequency of the same ultra-low noise reference derived from a
cryogenic sapphire oscillator. Practically, the reference signal at 11.98\,GHz is divided to generate a 100\,MHz reference which
is disseminated to FO1 and FOM. Being located in the neighboring lab, FO2 benefits from using the 11.98\,GHz directly. Another
1\,GHz reference is also generated from the 11.98\,GHz signal and sent through a fiber link as a reference for the fiber
femtosecond optical frequency comb. The 100\,MHz signal is also compared to the 100\,MHz output of a H-maser. A slow phase-locked
loop (time constant of 1000\,s) is implemented to ensure coherence between the reference signals and the H-maser to avoid long
term frequency drift of the reference signals.

During the 15 days of measurement reported here, the three fountains are operated continuously as primary frequency standards
measuring the same reference oscillator. The overall frequency instability for this measurement is $3.5\times 10^{-14}$ at 1\,s
for FO2, $4.2\times 10^{-14}$ at 1\,s for FO1 and $7.2\times 10^{-14}$ at 1\,s for FOM. The accuracy of these clocks are $4\times
10^{-16}$ for FO1 and FO2 and $1.2\times 10^{-15}$ for FOM. The fractional frequency differences between the fountain clocks are
all consistent with zero within the combined 1-sigma error bar, which implies consistency to better than $10^{-15}$.

The link between the cavity stabilized laser and the frequency comb is a fiber link of 50\,m length with a phase noise
cancellation system. The probe laser beam is sent to the atoms after passing through an acousto-optic modulator (AOM) driven at a
computer controlled frequency. Each transition is measured for 32 cycles before switching to the transition involving opposite
$m_F$ states. Two digital servo-loops to both atomic resonances therefore run in parallel with interlaced measurements. For each
servo-loop we alternately probe both sides of the resonance peak. The difference between two successive transition probability
measurements constitutes the error signal used to servo-control the AOM frequency to the atomic transition. In addition, we
interlace sets of 64 cycles involving two different trapping depths. The whole sequence is repeated for up to one hour.

This operating mode allows the independent evaluation of three clock parameters. The difference between the frequency
measurements made for each Zeeman sub-state can be used to accurately determine the magnetic field. As we switch to the other
resonance every 32 cycles, this gives a real-time calibration of the magnetic-field averaged over 64 cycles. The global average
of the measurement is the value of the clock frequency and is independent on the first order Zeeman effect as the two probed
transitions are symmetrically shifted. Finally, the two frequencies corresponding to two different dipole trap depths are used
for a real-time monitoring of the possible residual light shift of the clock transition by the optical lattice.

\begin{figure}
\begin{center}
\resizebox{0.75\columnwidth}{!}{
\includegraphics{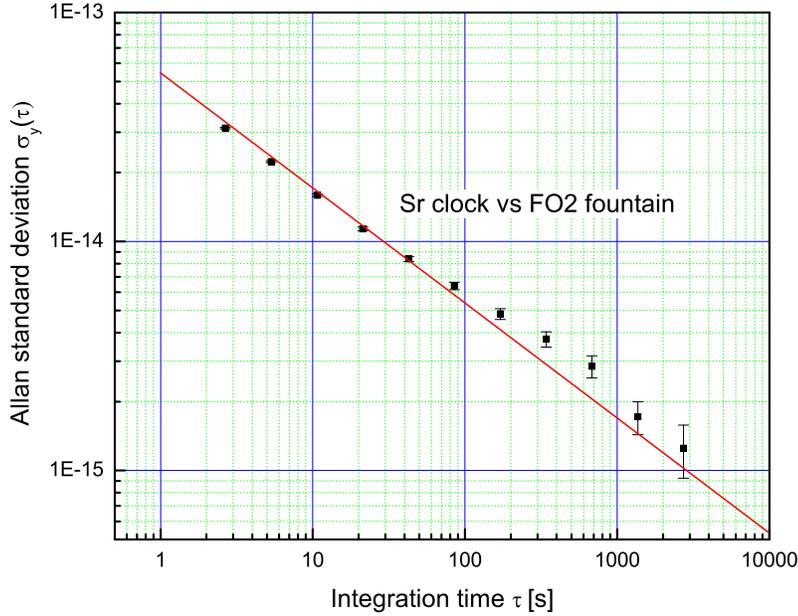} }
\caption{Allan standard deviation of the frequency measurements for a magnetic field $B=0.87$\,G and a time of interrogation of
20\,ms. The line is a fit to the data using a $\tau^{-1/2}$ law. The corresponding stability at 1\,s is $6\times 10^{-14}$.}
\label{fig:variance}
\end{center}
\end{figure}

The frequency stability of the Sr lattice clock-FO2 fountain comparison is shown in Fig.\,\ref{fig:variance}. The Allan deviation
is $6\times 10^{-14} \tau^{-1/2}$ so that the statistical uncertainty after one hour of averaging time is $10^{-15}$,
corresponding approximately to 0.5\,Hz.

\subsection{The frequency comb} \label{sec:femto}

For the absolute frequency measurement of the Sr transition we have used a fibre-based optical frequency comb which is based on a
FC1500 optical frequency synthesizer supplied by Menlo Systems. The laser source for the FC1500 comb is a passively mode-locked
femtosecond fibre laser which operates at a centre wavelength of approximately 1550\,nm and has a repetition rate of 100\,MHz.
The repetition rate can be tuned over approximately 400\,kHz, by means of an end mirror mounted on a translation stage controlled
by a stepper motor and a piezoelectric transducer. The output power from the mode-locked laser is split and fed to three
erbium-doped fiber amplifiers (EDFAs). These are used to generate three phase-coherent optical frequency combs whose spectral
properties can be independently optimized to perform different functions. The output from the first EDFA is broadened using a
nonlinear fibre to span the wavelength range from approximately 1000\,nm to 2100\,nm. This provides the octave-spanning spectrum
required for detection of the carrier-envelope offset frequency $f_0$ using the self-referencing
technique\,\cite{Jones00,Holzwarth00}. With proper adjustment of the polarisation controllers a beat signal with a signal to
noise ratio SNR$=40$\,dB in a resolution bandwidth of 100\,kHz is achieved. The electronics for the stabilization of the carrier
offset frequency comprises a photo detector, a tracking oscillator, and a digital phase-locked loop (PLL) with adjustable gain
and bandwidth. The offset frequency is stabilized by feedback to the pump laser diode current. The critical frequency to be
measured is the pulse repetition frequency $f_{rep}$ since the optical frequency is measured as a very high multiple of this
pulse repetition frequency. To overcome noise limitations due to the locking electronics and to enhance the resolution of the
counting system, we detect $f_{rep}$ at a high harmonic using a fast InGaAs photodiode with a bandwidth of 25\,GHz. Locking of
the repetition frequency is provided by an analogue phase-locked loop comparing the $90^{\mathrm{th}}$ harmonic of $f_{rep}$ with
a microwave reference and controlling the cavity length. The subsequent use of a harmonic tracking filter further enhances the
short term resolution of our counting system. For this purpose, the 9\,GHz beat signal is down-converted with a low noise 9\,GHz
signal synthesized from the microwave frequency reference (CSO / hydrogen maser referenced to a Cs-fountain clock). The
difference frequency is again multiplied by 128, reducing frequency counter digitization errors to below the level of the noise
of the microwave reference. Thereby, the frequency at which $f_{rep}$ is effectively measured is 1.15\,THz. A second EDFA
generates high power radiation which is frequency-doubled using a PPLN crystal, generating a narrow-band frequency comb around
780\,nm. This is subsequently broadened in a nonlinear fiber to generate a comb spanning the range 600-900\,nm. Light of the
spectroscopy laser at 698\,nm is superimposed on the output of the frequency comb using a beam splitter. To assure proper mode
matching of the beams and proper polarization adjustment one output of the beam splitter is launched into a single mode fiber and
detected with a DC photodetector. The other output is dispersed by means of a diffraction grating. A subsequent pinhole placed in
front of the photodetector then selects a narrow range of the spectrum at 698\,nm and improves the signal to noise ratio of the
observed heterodyne beat. For the heterodyne beat signal with the Sr clock laser a SNR of 30-35\,dB in a bandwidth of 100\,kHz
was achieved. Again, a tracking oscillator is used for optimal filtering and conditioning of the heterodyne signal. All beat
frequencies and relevant AOM-frequencies were counted using totalizing counters with no dead-time; the counters correspond to
$\Pi$-estimators\,\cite{Dawkins07} for the calculation of the standard Allan variance.

\section{First order Zeeman effect}

\begin{figure}
\begin{center}
\resizebox{0.75\columnwidth}{!}{
\includegraphics{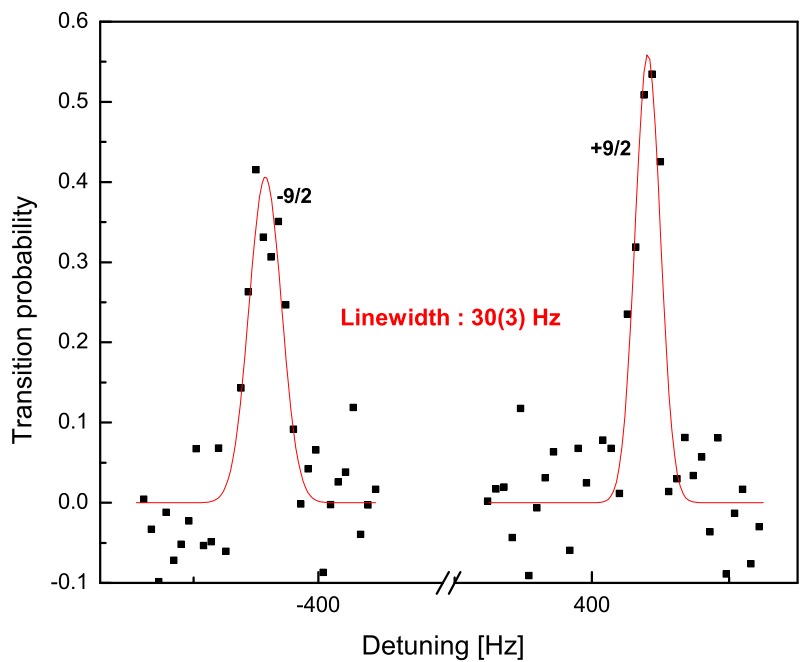} }
\caption{Experimental resonances observed for the $m_F=9/2\rightarrow m_F=9/2$ (left) and $m_F=-9/2\rightarrow m_F=-9/2$ (right)
transitions for a magnetic field $B=87\,\mu$T and an interrogation time of 20\,ms. The lines are gaussian fits to the data. The
asymmetry between both resonances results from the imperfection of the optical pumping.} \label{fig:resonances}
\end{center}
\end{figure}

In presence of a magnetic field, both clock levels, which have a total momentum $F=9/2$, are split into 10 Zeeman sub-states. The
linear shift $\Delta_Z$ of a sub-state due to a magnetic field $B$ is \begin{eqnarray} \Delta_Z=m_Fg_F\mu_BB/h,
\label{eq:zeemanshift}
\end{eqnarray}
where $m_F$ is the Zeeman sub-state (here $9/2$ or $-9/2$), $g_F$ the Land\'{e} factor of the considered state, $\mu_B$ the Bohr
magneton, and $h$ the Planck constant. Using the differential g-factor between $^3P_0$ and $^1S_0$ reported in
Ref.\,\cite{Boyd072}: $\Delta g=7.77(3)\times 10^{-5}$, we can determine the magnetic field by measuring two symmetrical
resonances. Fig.\,\ref{fig:resonances} shows the typical resonances observed with a magnetic field $B=87\,\mu$T for both $m_F=\pm
9/2$ sub-states. The linewidth is of the order of 30\,Hz, essentially limited by Fourier broadening hence facilitating the lock
of the frequency to each resonance. This linewidth corresponds to an atomic quality factor $Q=1.4\times 10^{13}$. At this
magnetic field, two successive $\pi$-transitions are separated by 96\,Hz, which is high enough to entirely resolve the Zeeman
sub-structure with that type of resonance and to limit possible line-pulling effects to below $10^{-15}$ (see
section\,\ref{sec:budget}).

The magnetic field used for pumping and detecting the atoms is provided by two coils in Helmoltz configuration to produce a
homogeneous field at the center of the trap. They are fed by a fast computer-controlled power supply to reach the desired value
in a few ms. This setup requires to accurately characterize the stability of the magnetic field, as the residual magnetic field
fluctuations are a possible issue for the clock accuracy and stability. The Zeeman effect can provide a precise measurement of
this field and its calibration when we measure the clock transition for two symmetrical transitions. When probing the two
transitions for the $m_F=\pm 9/2$ sub-states, the difference $\Delta\nu$ between the two frequencies can be related to the
magnetic field using Eq.\,\ref{eq:zeemanshift}: $\Delta\nu=9\Delta g\mu_BB/h$.

\begin{figure}
\begin{center}
\resizebox{0.75\columnwidth}{!}{
\includegraphics{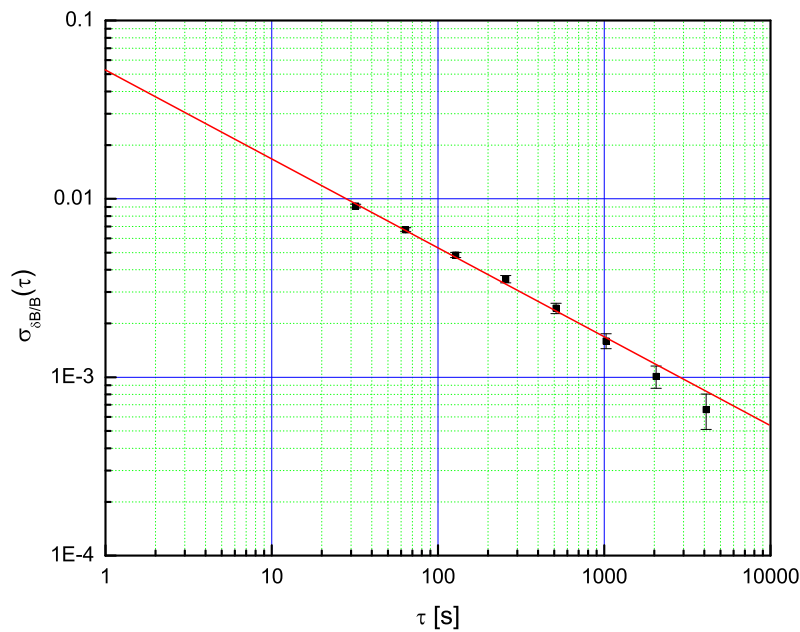} }
\caption{Allan standard deviation of the magnetic field. The line is a $\tau^{-1/2}$ fit to the data.}
\label{fig:magfieldstability}
\end{center}
\end{figure}

To evaluate the stability of the magnetic field, we chose a particular set of parameters ($B=87\,\mu$T and a modulation depth of
the numerical servo-loop of 10\,Hz) and repeated a large number of times the corresponding time sequence as described in
section\,\ref{sec:meas}. The measured magnetic field is averaged over 64 cycles. We then concatenated all the averaged data and
calculated the Allan standard deviation to determine the long term stability of the magnetic field. The result is plotted on
Fig.\,\ref{fig:magfieldstability}. The deviation is $10^{-2}$ in fractional units at 32\,s, and going down following a
$\tau^{-1/2}$ law for longer times. The deviation for long times is below $10^{-3}$. This measurement is totally dominated by the
frequency noise of the Sr clock and no fluctuations of the field itself are visible at the present level of resolution. For a
magnetic field of $87\,\mu$T, this represents a control of the magnetic field at the sub-$\mu$T level over long timescales.

\section{Frequency accuracy}

\subsection{Second order Zeeman effect} \label{sec:quadzeeman}

\begin{figure}
\begin{center}
\resizebox{0.75\columnwidth}{!}{
\includegraphics{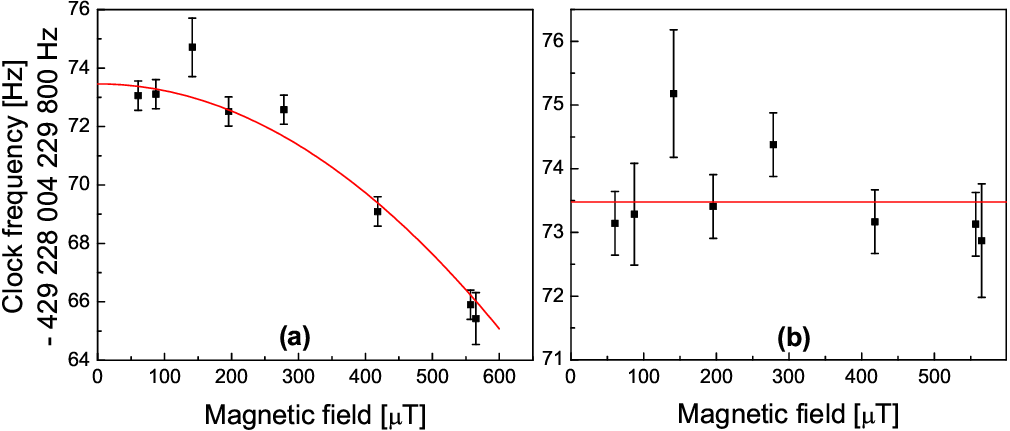} }
\caption{(a) Clock frequency as a function of the applied magnetic field. The line represents a fit of the experimental data by a
quadratic law with one adjustable parameter : the frequency offset. The linear term was set to 0 and the quadratic term to its
theoretical value. (b) Clock frequency after correction for the second order Zeeman effect. The line is the average of the data.}
\label{fig:quadzeeman}
\end{center}
\end{figure}

The clock has been operated with different values of the bias magnetic field up to $0.6$\,mT. As explained before, our method of
interrogation makes the measurements independent on the first order Zeeman effect. On the other hand, both resonances are shifted
by the same quadratic Zeeman shift which has to be evaluated. From the calibration of the magnetic field, we can evaluate the
dependence of the transition frequency as a function of this field. The results are plotted on Fig.\,\ref{fig:quadzeeman}. The
line plotted on the graph represents the expected quadratic dependence of $-23.3$\,Hz/mT$^2$\,\cite{Taichenachev062} where we
adjusted only the frequency offset to fit the data. The statistical uncertainty on this fit is 0.2\,Hz, and there is no
indication for a residual first order effect to within less than 1\,Hz at 0.6\,mT. At $B=87\,\mu$T, where most of the
measurements were done, the correction due to the quadratic Zeeman effect is 0.1\,Hz only. Conversely, an experimental value for
the quadratic Zeeman effect coefficient can be derived from the data plotted in Fig.\,\ref{fig:quadzeeman} with a 7\%
uncertainty. We find $-24.9(1.7)$\,Hz/mT$^2$, which is in agreement with theory.

\subsection{Residual lattice light shift} \label{sec:trap}

\begin{figure}
\begin{center}
\resizebox{0.75\columnwidth}{!}{
\includegraphics{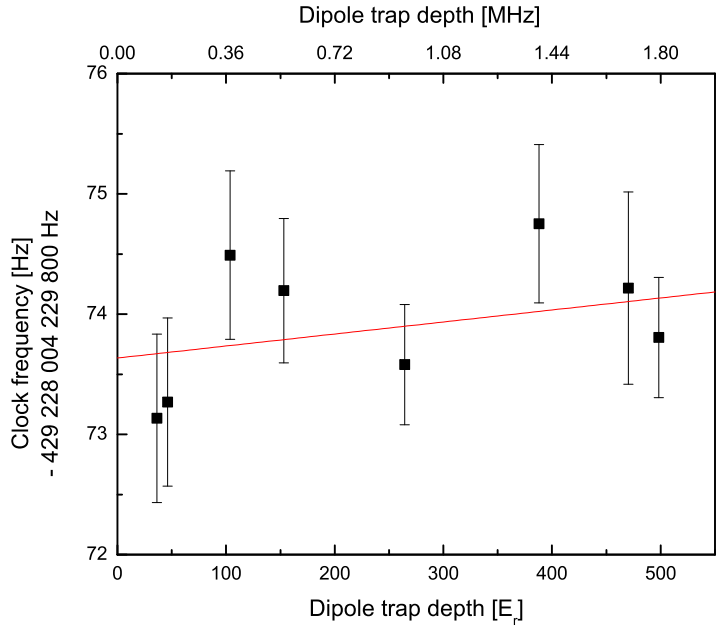} }
\caption{Clock frequency as a function of the dipole trap depth in terms of recoil energies. On the upper scale is the
corresponding light shift of the clock levels. The line is a linear fit to the data. The value of the light shift due to the trap
at 500\,$E_r$ is only 0.5(0.5)\,Hz.} \label{fig:trapeffect}
\end{center}
\end{figure}

The clock frequency as a function of the trapping depth is plotted on Fig.\,\ref{fig:trapeffect}. Measurements have been done
with depths ranging from 50 to 500\,$E_r$, corresponding to an individual light shift of both clock levels levels up to 1.8\,MHz.
Over this range, the scatter of points is less than 2\,Hz and the statistical uncertainty of each point lower than 1\,Hz. The
control of this effect has been evaluated by fitting the data with a line. The slope represents a shift of 0.5(5)\,Hz at
500\,$E_r$. The differential shift between both clock states is therefore controlled at a level of $3\times 10^{-7}$. In ultimate
operating conditions of the clock, a trapping depth of 10\,$E_r$ is theoretically sufficient to cancel the motional effects down
to below $10^{-17}$\,\cite{Lemonde05}. The light shift corresponding to this depth is 36\,kHz for both level, or $8\times
10^{-11}$ in fractional units. The kind of control demonstrated here would correspond, in these ultimate conditions, to a
residual light shift below $2\times 10^{-17}$.

\subsection{Uncertainty budget} \label{sec:budget}
Other systematic effects have been evaluated and included in the accuracy budget listed in Table\,\ref{tab:budget}.

The line pulling by neighbouring transitions has been carefully evaluated. Two types of transitions should be considered:
transverse motional sidebands and transitions between the various Zeeman states of the atoms.

Transverse motional sidebands can be excited by the transverse $k$-content of the probe laser. For a lattice depth of $100\,E_r$,
the transverse oscillation frequency is about 150\,Hz. Both diffraction and misalignement are below $1\,$mrad here so that the
transverse dynamics is deeply in the Lamb-Dicke regime and the height of transverse sidebands are expected to be at most $5\times
10^{-3}$ of the carrier (experimentally, they do not emerge from the noise of the measurements). The corresponding line pulling
is therefore below $0.4\,$Hz. This is confirmed by the absence of pathological dependence of the clock frequency as a function of
the lattice depth (Fig.\,\ref{fig:trapeffect}).

Unwanted Zeeman transitions result from the imperfection of the optical pumping process and of the polarization of the probe
laser. In standard configuration the only visible stray resonance is the $m_F=\pm7/2-m_F=\pm7/2$ transition, with a height that
is about half of the one of the $m_F=\pm9/2-m_F=\pm9/2$ resonance. It is difficult to set a realistic theoretical upper limit on
this effect since we have no direct access to the level of coherence between the various $m_F$ states, nor on the degree of
polarization of the probe laser. Experimentally however, several parameters can be varied to test the effect. The magnetic field
dependence shown in Fig.\,\ref{fig:quadzeeman} shows no deviation from the expected law to within the error bars. On the other
hand, measurements performed with various depths of the servo-loop modulation also show no differences to within $0.5\,$Hz.
Finally, we operated the clock with a probe laser polarization orthogonal to the bias field and using the $m_F=\pm9/2-m_F=\pm7/2$
transitions as clock resonances. The clock frequency in this configuration is found $0.2(5)$\,Hz away from the frequency in the
standard configuration. These measurements also test for a possible line pulling from higher order stray resonances involving
both a change of the transverse motion and of the internal Zeeman state which can fall very close to the main resonances.
Altogether we estimate a $0.5\,$Hz uncertainty on these line pulling effects.

Incidently, the combination of the measurements using $\sigma$ and $\pi$ transitions allows the derivation of the differential
Land\'{e} factor between both clock states\,\cite{Boyd072}. We find $g(^3P_0)-g(^1S_0)=7.90(7)\times 10^{-5}$, a value that
differs from the one reported in Ref.\,\cite{Boyd072} by twice the combined 1-sigma uncertainty.

The light shift due to the probe laser is essentially due to the off-resonant coupling of the $^3P_0$ state with the $^3S_1$
state. The typical light intensity used for the clock evaluation was of a few mW/cm$^2$. By varying this intensity by a factor up
to 2, no visible effect has been observed to within the uncertainty. A more precise evaluation was carried out using the bosonic
isotope $^{88}$Sr\,\cite{Baillard07}. The measured light shift for an intensity of $6$\,W/cm$^2$ was in this case of
$-78$(11)\,Hz. The corresponding effect for our current setup, where the probe power is 3 orders of magnitude smaller, is about
$0.1$\,Hz with an uncertainty in the $10^{-2}$\,Hz range.

The Blackbody radiation shift is derived from temperature measurements of the vacuum chamber using two Pt resistors placed on
opposite sides of the apparatus and using the accurate theoretical calculation reported in Ref.\,\cite{Porsev06}. The Blackbody
radiation shift in our operating conditions is $2.39(10)$\,Hz.

\begin{figure}
\begin{center}
\resizebox{0.75\columnwidth}{!}{
\includegraphics{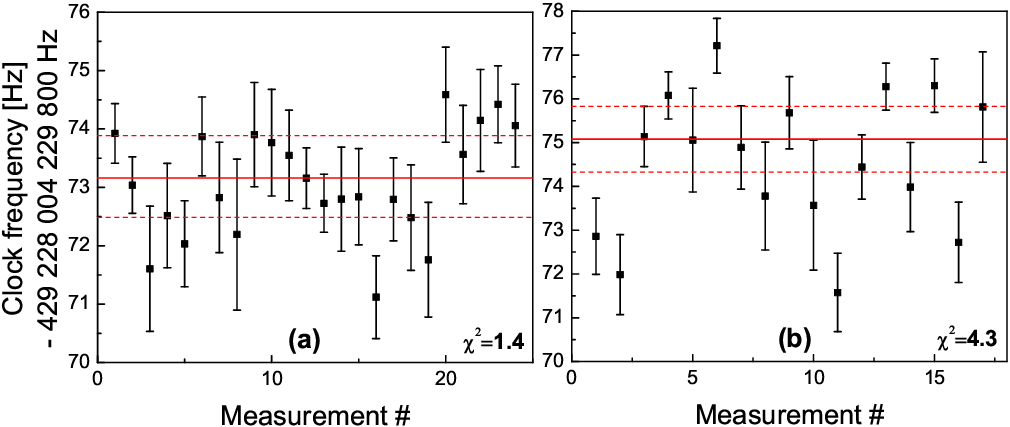} }
\end{center}
\caption{Two series of measurements performed with different clock parameters (see text). The series plotted on the right hand
side of the figure clearly exhibits a scatter of points that is incompatible with the individual statistical error bars of the
measurements. Its reduced $\chi^2$ is 4.3. The other series behaves normally and is shown for reference.}\label{fig:scatter}
\end{figure}

Finally a 1\,Hz uncertainty is attributed to an effect that has not been clearly identified. After having varied all the
parameters necessary for estimating the systematic effects, we decided to check the overall consistency by performing three
series of measurements with fixed parameters. The two first ones were performed with a bias field of $87\,\mu$T and servo loop
modulation depths of 7 and 10 Hz respectively. The third one with a larger field of $140\,\mu$T and a modulation depth of 7 Hz.
The results of series 2 and 3 are shown in Fig.\,\ref{fig:scatter}, where the error bars include the statistical uncertainty of
each measurement only. The scatter of points of series 3 is clearly incompatible with the individual error bars (the reduced
$\chi^2$ of this distribution is 4.3). In addition, its average value is 1.5\,Hz away. Having not clearly identified the reason
for this behaviour (one possibility could be a problem in the injection locking of one of the slave lasers at 698\,nm), we
decided to keep this series of data. We also cannot ensure that the effect is not present (though at a smaller level) in the
other measurements and decided to attribute an uncertainty of 1\,Hz to this effect.

Taking into account these systematic effects, the averaged clock frequency is determined to be
$\nu_{clock}=$\,429\,228\,004\,229\,873.6(1.1)\,Hz. The global uncertainty, $2.6\times 10^{-15}$ in fractional units, corresponds
to the quadratic sum of all the uncertainties of the systematic effects listed in Table\,\ref{tab:budget}. The statistical
uncertainty is at the level of 0.1\,Hz.

\begin{table}
\caption{Uncertainty budget.} \label{tab:budget}
\begin{tabular}{llll}
\hline\noalign{\smallskip}
Effect & Correction (Hz) & Uncertainty (Hz) & Fractional uncertainty ($\times 10^{-15}$)  \\
\noalign{\smallskip}\hline\noalign{\smallskip}
Zeeman & 0.1 & 0.1 & 0.2 \\
Probe laser Stark shift & 0.1 & $<0.1$ & $<0.1$ \\
Lattice AC Stark shift (100\,$E_r$) & 0 & 0.2 & 0.4 \\
Lattice 2nd order Stark shift (100\,$E_r$) & 0 & 0.1 & 0.2 \\
Line pulling (transverse sidebands) & 0 & 0.5 & 1.1 \\
Cold collisions & 0 & 0.1 & 0.2 \\
Blackbody radiation shift & 2.39 & $0.1$ & $0.1$ \\
See text & 0 & 1 & 2.3 \\
Fountain accuracy & 0 & 0.2 & 0.4 \\
\noalign{\smallskip}\hline\noalign{\smallskip}
Total & 2.59 & 1.1 & 2.6 \\
\noalign{\smallskip}\hline
\end{tabular}
\end{table}

\section{Conclusion} \label{sec:conclusion}

We have reported here a new measurement of the frequency of the $^1S_0\rightarrow\,^3P_0$ transition of $^{87}$Sr with an
uncertainty of 1.1\,Hz or $2.6\times 10^{-15}$ in fractional units. The result is in excellent agreement with the values reported
by the JILA group with a similar uncertainty\,\cite{Boyd07} and by the Tokyo group with a 4\,Hz error bar\,\cite{Takamoto06}.
Obtained in independent experiments with significant differences in their implementation, this multiple redundancy strengthens
the obtained results and further confirms the possibility to build high accuracy clocks with cold atoms confined in an optical
lattice. It also further assesses this transition as a possible candidate for a future redefinition of the second.

SYRTE is Unit\'{e} Associ\'{e}e au CNRS (UMR 8630) and a member of IFRAF. This work is supported by CNES and DGA. PTB
acknowledges financial support from the German Science foundation through SFB 407.

\bibliographystyle{epj}

\begin{thebibliography}{21}

\bibitem{Boyd07}
M.M. Boyd, A.D. Ludlow, S.~Blatt, S.M. Foreman, T.~Ido, T.~Zelevinsky, J.~Ye,
  Phys. Rev. Lett. \textbf{98}, 083002 (2007)

\bibitem{KatoPal03}
H.~Katori, M.~Takamoto, V.G. Pal'chikov, V.D. Ovsiannikov, Phys. Rev. Lett.
  \textbf{91}, 173005 (2003)

\bibitem{Takamoto03}
M.~Takamoto, H.~Katori, Phys. Rev. Lett. \textbf{91}(22), 223001 (2003)

\bibitem{Ludlow06}
A.D. Ludlow, M.M. Boyd, T.~Zelevinsky, S.M. Foreman, S.~Blatt, M.~Notcutt,
  T.~Ido, J.~Ye, Phys. Rev. Lett. \textbf{96}, 033003 (2006)

\bibitem{Brusch06}
A.~Brusch, R.~{Le Targat}, X.~Baillard, M.~Fouch{\'e}, P.~Lemonde, Phys. Rev.
  Lett. \textbf{96}, 103003 (2006)

\bibitem{Boyd06}
M.M. Boyd, T.~Zelevinsky, A.D. Ludlow, S.M. Foreman, S.~Blatt, T.~Ido, J.~Ye,
  Science \textbf{314}, 1430 (2006)

\bibitem{Lemonde05}
P.~Lemonde, P.~Wolf, Phys. Rev. A \textbf{72}, 033409 (2005)

\bibitem{Courtillot03}
I.~Courtillot, A.~Quessada, R.P. Kovacich, A.~Brusch, D.~Kolker, J.J. Zondy,
  G.D. Rovera, P.~Lemonde, Phys. Rev. A \textbf{68}, 030501(R) (2003)

\bibitem{Takamoto06}
M.~Takamoto, F.L. Hong, R.~Higashi, Y.~Fujii, M.~Imae, H.~Katori, J. Phys. Soc.
  Jpn. \textbf{75}, 104302 (2006)

\bibitem{Letargat06}
R.~{Le Targat}, X.~Baillard, M.~{Fouch\'e}, A.~Brusch, O.~Tcherbakoff, G.D.
  Rovera, P.~Lemonde, Phys. Rev. Lett. \textbf{97}, 130801 (2006)

\bibitem{Wolf063}
P.~Wolf, G.~Petit, E.~Peik, C.~Tamm, H.~Schnatz, B.~Lipphardt, S.~Weyers,
  R.~Wynands, J.Y. Richard, S.~Bize et~al., \emph{Comparing high accuracy
  frequency standards via TAI}, in \emph{Proc. of $20^{th}$ European Frequency
  and Time Forum} (Braunschweig, Germany, 2006)

\bibitem{Baillard06}
X.~Baillard, A.~Gauguet, S.~Bize, P.~Lemonde, P.~Laurent, A.~Clairon,
  P.~Rosenbusch, Opt. Comm. \textbf{266}, 609 (2006)

\bibitem{Tamura93}
K.~Tamura, J.~Jacobson, E.P. Ippen, H.A. Haus, J.G. Fujoimoto, Opt. Lett.
  \textbf{18}, 220 (1993)

\bibitem{Kubina05}
P.~Kubina, P.~Adel, F.~Adler, G.~Grosche, T.~Hänsch, R.~Holzwarth,
  A.~Leitenstorfer, B.~Lipphardt, H.~Schnatz, Opt. Express \textbf{13}, 904
  (2005)

\bibitem{Bize04}
S.~Bize, P.~Laurent, M.~Abgrall, H.~Marion, I.~Maksimovic, L.~Cacciapuoti,
  J.~Gr\"{u}nert, C.~Vian, F.~{Pereira dos Santos}, P.~Rosenbusch et~al., C. R.
  Physique \textbf{5}, 829 (2004)

\bibitem{Bize05}
S.~Bize, P.~Laurent, M.~Abgrall, H.~Marion, I.~Maksimovic, L.~Cacciapuoti,
  J.~Gr\"{u}nert, C.~Vian, F.~{Pereira dos Santos}, P.~Rosenbusch et~al., J.
  Phys. B: At. Mol. Opt. Phys. \textbf{38}, S449 (2005)

\bibitem{Jones00}
D.J. Jones, S.A. Diddams, J.K. Ranka, A.~Stentz, R.S. Windeler, J.L. Hall, S.T.
  Cundiff, Science \textbf{288}, 635 (2000)

\bibitem{Holzwarth00}
R.~Holzwarth, T.~Udem, T.W. {H{\"a}nsch}, J.C. Knight, W.J. Wadsworth, P.S.J.
  Russel, Phys. Rev. Lett. \textbf{85}, 2264 (2000)

\bibitem{Dawkins07}
S.T. Dawkins, J.J. Mcferran, A.N. Luiten, IEEE Trans. Ultrason., Ferroelect.,
  Freq. Contr. \textbf{54}, 918 (2007)

\bibitem{Boyd072}
M.M. Boyd, T. Zelevinsky, A.D. Ludlow, S.~Blatt, T.~{Zanon-Willette}, S.M. Foreman,
  J.~Ye, Phys. Rev. A \textbf{76}, 022510 (2007)

\bibitem{Taichenachev062}
A.V. Taichenachev, V.I. Yudin, C.W. Oates, C.W. Hoyt, Z.W. Barber, L.~Hollberg,
  Phys. Rev. Lett. \textbf{96}, 083001 (2006)

\bibitem{Baillard07}
X.~Baillard, M.~Fouch\'{e}, R.~{Le Targat}, P.G. Westergaard, A.~Lecallier,
  Y.~Lecoq, G.D. Rovera, S.~Bize, P.~Lemonde, Opt. Lett. \textbf{232}, 1812
  (2007)

\bibitem{Porsev06}
S.G. Porsev, A.~Derevianko, Phys. Rev. A \textbf{74}, 020502 (2006)

\end{thebibliography}

\end{document}